\documentclass[prl,twocolumn,superscriptaddress,showpacs,floatfix,amsfonts]{revtex4}
\usepackage{graphicx,graphics,color,epsfig}
\usepackage{bm}
\usepackage{amsmath}
\usepackage{amssymb}
\usepackage{epstopdf}
\begin{document}
\preprint{}

\title{Strong Correlations and Magnetic Frustration in the High $T_c$ Iron Pnictides}
\author{Qimiao Si}
\affiliation
{Department of Physics \& Astronomy, Rice University, Houston, 
TX 77005}
\author{Elihu Abrahams}
\affiliation
{Center for Materials Theory, Serin Physics Laboratory, Rutgers University, Piscataway, New Jersey 08855}
\date{\today}
\begin{abstract}
We consider the iron pnictides in terms 
of a proximity
to a Mott insulator. The superexchange
interactions contain competing nearest-neighbor and 
next-nearest-neighbor components. 
In the undoped parent compound, these frustrated interactions
lead to a two-sublattice collinear antiferromagnet 
(each sublattice forming a N\'{e}el ordering),
with a reduced magnitude for the ordered moment.
Electron or hole doping, together
with the frustration effect,
suppresses the magnetic ordering and allows a superconducting 
state. The exchange interactions favor a d-wave 
superconducting order parameter; in the notation
appropriate for the Fe square lattice, its orbital
symmetry is $d_{xy}$.
A number of existing and future experiments are discussed
in light of the theoretical considerations.

\end{abstract}


\maketitle

{\it Introduction:~~}
High $T_c$ superconductivity has recently been discovered
in the iron pnictides, with the F-doped LaOFeAs being 
the prototype \cite{Kamihara_FeAs}.
Variations include P replacement for As, Ni replacement for Fe,
and 
rare-earth replacements for La \cite{Kamihara_FeP,Chen,Wang,Zhao,Wen}.
In addition, F substitution
for O, which adds itinerant electron carriers to the system, 
could be replaced by, {\it e.g.}, Sr substitution for La, which 
introduces hole doping \cite{hole_doped}.

Like the cuprates, 
the iron pnictides
have a layered structure.
The FeAs unit
appears to contain all the electronic states near the
Fermi energy, similar to the case of the CuO$_2$ layer in the cuprates.
The electrons partially occupying the $d$-orbitals of the
iron sites can be strongly correlated, as are those
on the copper sites in the cuprates.
At the same time, there are also important differences between 
the two classes of materials.
In this letter, we consider the consequences of the unique 
aspects of the electronic states of the iron pnictides.
We will frame our discussion in terms of the
F-doped LaOFeAs family, and touch upon their 
cousin compounds where appropriate.

\begin{figure}[th]
\vskip 1 cm
\centerline{\psfig{file=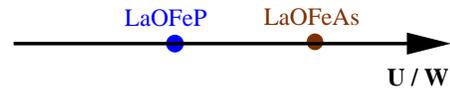, width=6cm}}
\vskip 0.5 cm
 \caption{Placing ${\rm LaOFeAs}$ and ${\rm LaOFeP}$ in terms 
of the control parameter $U/W$, where $U$ is the on-site Coulomb
interaction and $W$ the effective bandwidth.
} \label{U-over-W}
\end{figure}

One basic question is whether the 
Mott insulating physics plays
any significant role in ${\rm LaOFeAs}$.
The answer is 
not 
necessarily affirmative;
there are, for instance, indications from bandstructure
calculations that covalency is sizable
not only in 
${\rm LaOFeP}$ \cite{Lebegue}
but also in ${\rm LaOFeAs}$ \cite{Singh}.
Nonetheless, we argue that there are indirect evidences for the case that ${\rm LaOFeAs}$ is in proximity
to a Mott insulator.
First, the measured electrical resistivity is very large,
$\rho \approx 
5 {\rm m \Omega cm} $ at room temperature \cite{Kamihara_FeAs}.
This corresponds
to a normalized
mean free path $k_F\ell
\approx h c / e^2 \rho \approx 0.5$
(where $c \approx 8.7 \AA$ is the lattice constant along the
normal to the FeAs plane, and 
$h/e^2 \approx 26 {\rm k \Omega}$ is the quantum
resistance), which qualifies the system
as a bad metal.
Second, ${\rm LaOFeP}$, which has a smaller lattice constant
($c=8.5\AA$), thus 
a larger internal pressure,
is expected to have 
a larger effective bandwidth $W$ but a similar  
effective Coulomb interaction $U$ compared to
${\rm LaOFeAs}$.
That ${\rm LaOFeP}$ has a smaller
$U/W$ 
compared to 
${\rm LaOFeAs}$ is
consistent with the observation that
${\rm LaOFeP}$ is a better metal; 
its $k_F\ell \approx 1 $ at the room temperature 
and, indeed, it is superconducting with 
$T_c \approx 4$~K \cite{Kamihara_FeP}.
These considerations are illustrated in 
Fig.~\ref{U-over-W}.
Additional evidence along this direction 
is provided by the lack of a Drude peak in both the 
measured optical conductivity of 
${\rm LaOFeAs}$ \cite{Dong}, 
as well as the calculated one by 
the DMFT+DFT method \cite{Haule}. Thus, our approach to ${\rm LaOFeAs}$ is motivated by important experimental observations: the very large resistivity and the absence of a Drude peak. These are unlike a Fermi liquid and therefore are not likely to be accounted for without including correlation effects. They imply that most of the electronic excitations lie in the incoherent part of the spectrum.
Even if ${\rm LaOFeAs}$
is not fully Mott insulating, it shouldn't be far
away from it. We therefore find it instructive
to discuss these systems from a strong-coupling point of view.

{\it Why is the magnetism so weak?~~}
Within this strong coupling framework, 
the issue arises as to why the magnetism is so weak,
both in the undoped and lightly-doped iron
pnictides.
Consider 
first the undoped parent compound. 
Valence counting 
in ${\rm LaOFeAs}$ yields 
${\rm Fe^{2+}}$, which contains six outermost-shell electrons
partially filling the five 3$d$ orbitals.
The degeneracy of the latter is split by the crystal field.

One characteristic feature 
seen in the
{\it ab initio} electronic structure 
calculations is that 
the splitting among the five 3$d$ orbitals is relatively small.
Ref.~\cite{Hirschfeld}
shows that the individual
separations among the $d$ levels is on the order of, or less than,
0.1 eV. 
Taking into account the typical Coulomb interactions $U$, of the
order of 4-5~eV and the Hund's coupling $J_H$, of the order of 
0.7~eV \cite{Haule}, we expect the six outermost-shell
electrons to occupy the 3$d$ orbitals in the scheme depicted
in Fig.~2a. The associated Mott
insulator has spin $S=2$.
The spin-orbit coupling is expected to be considerably
weaker than $J_H$, so we shall focus on the spin magnetism.
Even if the separations between the crystal levels were larger
than $J_H$ (but still smaller than $U$), 
there will still be 
a double degeneracy \cite{Li,Baskaran},
leading in our consideration to an $S=1$ Mott insulator.

Such a large-spin Mott insulator is expected to be strongly
magnetic. Yet, a
neutron scattering experiment \cite{Dai}
has shown that ${\rm LaOFeAs}$ is an antiferromagnet with a rather 
small ordered moment, on the order of $0.4\mu_B/$Fe.

\begin{figure}[th]
\vskip 1 cm
\centerline{\psfig{file=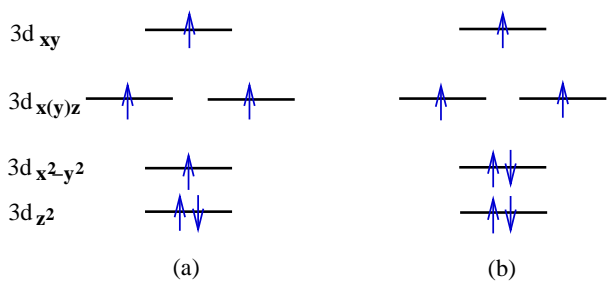, width=8cm}}
\vskip 0.5 cm
 \caption{(a) Spin-2 states 
relevant for the undoped LaOFeAs.
The crystal levels are according to Ref.~\cite{Hirschfeld}.
The $x$ and $y$ we use differ from the 
standard notation, adopted there, by a rotation
of 45$^o$; see the main text.
(b) Spin-3/2 states that become important when electron doping 
is introduced into the FeAs layer. Hole doping
will lead to the analogous spin-3/2 states, corresponding to
five electrons residing on the 3d orbitals.
} \label{Spin-states}
\end{figure}

The issue is even more acute in the doped cases.
Electron doping will introduce additional
states 
with spin 3/2, as illustrated in Fig.~2b.
Since the relevant local states are all magnetic,
it is surprising that a relatively small amount
of doping (say $10\%$ F-doping for O) does not preserve
the magnetic ordering.

{\it Magnetic frustration:~~}
We propose that the answer to these questions lies primarily
in
magnetic frustration. The key feature here is that
in the FeAs unit, not only do the Fe atoms form a square lattice,
but each As atom, away from the Fe plane, lies an equal
distance from each of the four adjacent Fe atoms. Because our
focus will be on the Fe plane, we find it convenient
to use the symmetry classification appropriate for the Fe
square lattice: we choose the $x$ and $y$ axes 
to be along the Fe-Fe bond direction; these are rotated by 45$^{\rm o}$ 
from the notation used in recent papers.

Consider the superexchange interactions between the
$3d_{x^2-y^2}$ orbitals of nearby Fe sites.
Inspection of the orbitals suggest that the 
strongest channel of hybridization will be with the 
As $4p_{x-y}$ or $4p_{x+y}$ orbital. 

For a pair of next-nearest-neighbor
(n.n.n.) Fe 
$3d_{x^2-y^2}$
spins, the lowest-energy
intermediate state mediating the superexchange 
interaction corresponds to two 
electrons occupying the same $4p_{x-y}$
(or $4p_{x+y}$) orbital (Fig.~3b). Since this intermediate
state is a singlet state, the resulting exchange 
interaction is {\it antiferromagnetic}:
\begin{eqnarray}
J_2 \approx 2 \frac{V_{x^2-y^2}^4}{(\epsilon_{p_{x-y}} 
- \epsilon_{d_{x^2-y^2}})^3} ,
\label{J2}
\end{eqnarray}
where $V_{x^2-y^2}$ is the hybridization matrix between
the Fe $3d_{x^2-y^2}$ and As $4p_{x-y}$ orbitals.

For a pair of  
nearest-neighbor (n.n.) $3d_{x^2-y^2}$
Fe spins, the lowest-energy intermediate
states correspond instead two electrons 
occupying a pair of 
distinct $4p_{x+y}$ and $4p_{x-y}$ orbitals (Fig.~3c).
The resulting exchange interaction is ferromagnetic:
\begin{eqnarray}
J_1 \approx - 2 V_{x^2-y^2}^4 \left [ \frac{1}{(\epsilon_{p_{x-y}} - 
\epsilon_{d_{x^2-y^2}}+J_{H,p})^3}\right . \nonumber \\
 - \left . \frac{1}{(\epsilon_{p_{x-y}} - 
\epsilon_{d_{x^2-y^2}})^3}\right ]. 
\label{J1}
\end{eqnarray}
Here $J_{H,p}<0$ is the 
Hund's coupling between the 
As 4$p_{x-y}$ and 4$p_{x+y}$ orbitals,
which favors the triplet intermediate state
over the singlet one. 
Notice that $J_{H,p}$ is relatively small, we expect 
that the ferromagnetic term is small compared to the 
antiferromagnetic one.

\begin{figure}[th]
\vskip 1 cm
\centerline{\psfig{file=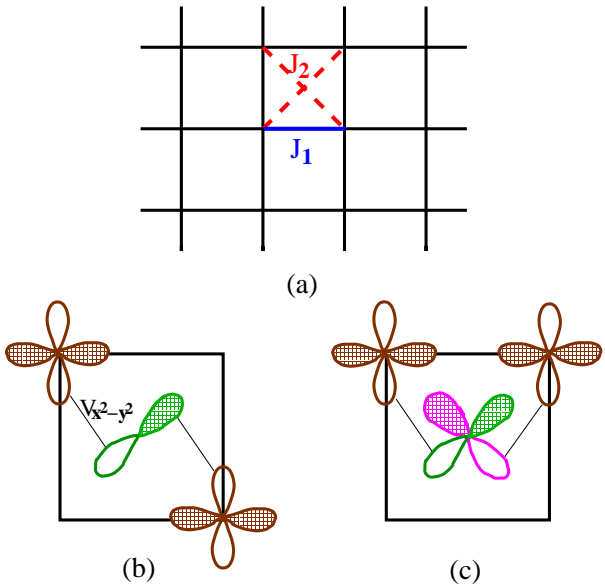, width=8cm}}
\vskip 0.5 cm
 \caption{(Color online) (a) The dominant superexchange interactions 
are those between both 
a pair of nearest-neighbor
Fe spins ($J_1$) and a pair of next-nearest-neighbor
Fe spins ($J_2$);
(b) The process contributing to the 
n.n.n. superexchange interaction between a pair of 
3$d_{x^2-y^2}$ electrons. It involves the same 
$4p_{x-y}$ orbital,
leading to an antiferromagnetic $J_2$;
(c) The process contributing to the 
n.n. superexchange interaction between a pair of 
3$d_{x^2-y^2}$ electrons. It involves two orthogonal 
4$p$ orbitals (green and magenta, respectively),
leading to a ferromagnetic $J_1$.
} \label{superexchange}
\end{figure}

The dominating matrix elements of the hybridization matrix
have been given in Ref.~\cite{Hirschfeld}.
The other relevant hybridizations involve 
the $3d_{x'z}$ and $3d_{y'z}$ orbitals.
(Here prime denotes the crystallographic axes, which are 
rotated by 45$^{o}$ from the Fe-square-lattice axes.)
For the n.n.n. interaction,
the dominating antiferromagnetic terms (with appropriate
replacements of the hybridization matrix elements in
Eq.~(\ref{J2})) appear in
the diagonal matrix elements 
$x'z-x'z$ and $y'z - y'z$, as well as in the 
off-diagonal matrix elements
$x'(y')z-(x^2-y^2)$.
For the n.n. exchange $J_1$, the dominating 
antiferromagnetic terms also appear in the 
diagonal matrix elements $x'z - x'z$ and 
$y'z - y'z$ and in the off-diagonal matrix elements
$x'(y')z-(x^2-y^2)$. Finally, the n.n.n. exchange $J_2$
involves virtual processes associated with only one As atom, 
while the n.n. exchange $J_1$ picks up contributions
from two As atoms.
Taken together, we expect that the largest eigenvalues 
of both the $J_2$ matrix and $J_1$ matrix correspond to
antiferromagnetic interactions, with the former larger 
than half of the latter.

The result is the following general form for the spin
Hamiltonian,
\begin{eqnarray}
H_J = \sum_{ij} J_{ij}^{\alpha\beta}
{\bf s}_{i,\alpha} \cdot {\bf s}_{j,\beta}
+J_H \sum_{i,\alpha \ne \beta} {\bf s}_{i,\alpha} \cdot {\bf s}_{i,\beta} ,
\label{H_J}
\end{eqnarray}
with mixed $J_{n.n}^{\alpha\beta}=J_1^{\alpha\beta}$  
but antiferromagnetic $J_{n.n.n.}=J_2^{\alpha\beta}$.
Here $J_1$ and $J_2$ are both matrices in the orbital
basis, with matrix elements labeled by $\alpha\beta$.
Again, whether the local states are spin 2 or spin 1,
corresponding to $\alpha$ or $\beta=1,2,3,4$ or 
$\alpha$ or $\beta=1,2$, depend on whether the Hund's coupling 
$J_H$ is large or small compared to the crystal level splittings.

Eq.~(\ref{H_J}) specifies a frustrated spin system.
In the 
$J_2>|J_1|/2$ case here,
the ground state
is expected to be 
a two-sublattice collinear antiferromagnet (with
each sublattice itself forming a 
N\'{e}el ordering) \cite{Chandra,Shannon}.
This spin pattern was first proposed 
for LaOFeAs
based on a consideration of the 
Fermi-surface nesting within a spin-density-wave
picture \cite{Dong,Mazin,Kuroki}.
It has subsequently been shown to be consistent with the 
elastic neutron scattering experiment in ${\rm LaOFeAs}$ \cite{Dai}.

Frustration effects are also important to yield an ordered moment 
that is considerably smaller than the atomic value of order
2$\mu_B$. The important point here is that it introduces $J_2/J_1$ 
as a tuning parameter, which allows for a reduction of the ordered
moment. Indeed, as $J_2/J_1$ is decreased towards a critical value,
the ordered moment is reduced to zero. The experimental value of the moment, as already mentioned, is about $0.4\mu_B/$Fe in ${\rm LaOFeAs}$ \cite{Dai}.

According to Ref.~\cite{Hirschfeld},
the $3d-4p_x$ hybridization matrix elements and the corresponding 
energy level separations are of the order of 0.8~eV and 1.3~eV, respectively.
The perturbative expression, Eq. (1), leads to a n.n.n. exchange coupling
$J_2$ of the order of 0.5~eV. While it is not expected 
to be quantitatively accurate, the result
does suggest that the exchange interaction will be sizable.

For the doped case, 
the effective model is
a matrix $t-J_1-J_2$ Hamiltonian:
\begin{eqnarray}
H_{tJ} = H_t + H_J .
\label{H_tJ}
\end{eqnarray}
The kinetic component of the Hamiltonian is 
\begin{eqnarray}
H_t = \sum_{ij} t_{ij}^{\alpha\beta}
{\tilde{c}}_{i,\alpha}^{\dag} {\tilde{c}}_{j,\beta} .
\label{H_t}
\end{eqnarray}
Here, the ${\tilde{c}}_{\alpha,i}$ describe 
constrained fermions, which connect the spin 2 and 
spin 3/2 configurations at the site $i$,
while $t_{n.n}^{\alpha\beta}=t_1^{\alpha\beta}$  
and $t_{n.n.n.}=t_2^{\alpha\beta}$ are the n.n.
and n.n.n. hybridization matrices.
(Recall that $\alpha,\beta$ refer to the $d$ orbitals.)
The net result of 
$H_t$ is to introduce transitions between the 
spin-2 and spin-3/2 states 
of the n.n. and n.n.n. Fe sites. 

Because the frustration in the superexchange interactions has
already reduced the ordered moment in the undoped parent 
compound, the magnetic ordering can be readily suppressed in
the doped materials. This further suppression occurs because 
the F-doping for O induces a spin-3/2 (or spin-1/2) substitution
of the spin-2 (or spin-1) states.
Experimentally, the absence of magnetic ordering 
has been shown in
${\rm LaO_{1-x}F_xFeAs}$, with electron doping of
${\rm x} \approx 8\%$ \cite{Dai}.

{\it Superconductivity:~~}
Frustration effects, while suppressing the magnetic ordering, accumulate
entropy at low temperatures. The relief of this entropy can take the form
of creating a superconducting order.
Precisely how this happens is one of the challenging questions
in strongly correlated electron systems.
Still, there are some general considerations we can make on the 
superconductivity.

The proximity to a Mott insulator disfavors isotropic s-wave order 
parameter for the superconducting state. Given that the n.n.n.
antiferromagnetic exchange interaction plays a dominant role in
the magnetic ordering at half-filling, it is natural that the
superconducting state has a $d_{xy}$ orbital symmetry.
(A mixed pairing state ($d_{x^2-y^2}+i d_{xy}$)
may also appear for certain range of $J_2/J_1$ \cite{Sachdev}.)
To see this explicitly, we carry through a simplified analysis
on a square plaquette. Our consideration parallels that
of Ref.~\cite{Scalapino_Trugman} 
for the cuprate case.

\begin{figure}[th]
\vskip 1 cm
\centerline{\psfig{file=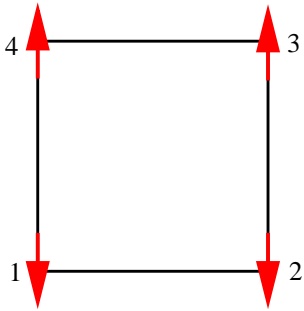, width=4cm}}
\vskip 0.5cm
 \caption{A square plaquette showing a spin arrangement of the two-sublattice
collinear antiferromagnet.
} \label{plaquette}
\end{figure}

The square plaquette we consider is illustrated in Fig.~\ref{plaquette}.
We will start from the spin-1/2 case.
The four-electron ($N=4$) ground state is
well approximated by 
\begin{eqnarray}
|N=4,gs\rangle  \propto (c_{4,\uparrow}^{\dagger}
c_{3,\uparrow}^{\dagger} 
c_{2,\downarrow}^{\dagger} 
c_{1,\downarrow}^{\dagger}  
+ S.R.) |0 \rangle
\label{gs_N=4}
\end{eqnarray}
where $|0\rangle$ is the vacuum state, and $S.R.$ denotes spin-reversal.

The $N=2$ ground state, at the same time, is well approximated
by a Gutzwiller-projected Slater determinant of two electrons,
which can be written in real space as 
\begin{eqnarray}
|N=2,gs \rangle \propto P_G \sum_{i,i'=1}^4 
c_{i,\downarrow}^{\dagger}
c_{i',\uparrow}^{\dagger} |0\rangle ,
\label{gs_N=2}
\end{eqnarray}
where $P_G$ eliminates any double occupancy of a site.
It is straightforward to show that the pairing operator 
that has the maximum matrix element between
$|N=4,gs\rangle$ 
and $|N=2,gs\rangle$ is the $d_{xy}$ pairing operator,
\begin{eqnarray}
\Psi_{xy} \propto \sum_{{\bf k}}
(\sin k_x \sin k_y) c_{{\bf k},\uparrow}
c_{-{\bf k},\downarrow}
\label{d_xy}
\end{eqnarray}
The construction of the exact 
$|N=4,gs\rangle$ and $|N=2,gs\rangle$ for
the spin=1/2 case, as well as 
the equivalent calculations for the 
higher spin cases, can be readily
done numerically.
The above suggests that
the magnetic exchange interactions 
will promote the $d_{xy}$ pairing,
regardless of the specific 
mechanism with which the exchange 
interactions cause superconductivity.

{\it Conclusion:~~} A complete analysis of the matrix $t-J_1-J_2$ model that we have introduced could reveal the existence of other competing phases, such as inhomogeneous magnetic structures. The description of the phase diagram and its evolution with doping are thus interesting subjects. Such studies must await more accurate determinations from {\it ab initio} calculations and/or experiment of all the underlying matrix elements [{\it e.g.} Eqs.\ (1,2)] and are beyond the scope of this letter.  

A key test for our picture is to experimentally determine
the relevant spin states in both the parent and doped systems,
as well as to measure the exchange interactions from, say,
the spin-wave spectra in the undoped parent compounds.
Studying additional families of materials which, in the undoped case,
can be placed along the $U/W$ axis (Fig.~\ref{U-over-W}),
will allow a fuller exploration
of the half-filled phase diagram and its doped counterpart.
This is especially important for 
the parts of the phase diagram that are either
deep inside the Mott insulating phase, or well into
the large-$k_F\ell$ non-superconducting metallic regime.

Immediately before
this paper was 
finalized,
we learnt of the work of T. Yildirim \cite{Yildirim},
which independently considered the frustration effect
using {\it ab initio} calculations, and which, in contrast
to many other density functional-based calculations, gave 
an ordered moment for the parent compound in the experimental
range. While we believe that a consideration of correlation
effects is essential to account for the ``bad metal" properties
described in the Introduction, there could be other
routes to the existence of a small ordered moment itself.

We thank E. Morosan, C. Broholm, H.-P. Cheng, P. Dai, K. Haule, 
P. Hirschfeld, G. Kotliar for useful discussions, and 
acknowledge the support partially provided 
by the NSF Grant No. DMR-0706625 and the Robert
A. Welch Foundation.

\end{document}